\documentclass[aps,preprint]{revtex4}%
\usepackage{amsfonts}%
\usepackage{amsmath}%
\usepackage{amssymb}%
\usepackage{graphicx}
\providecommand{\U}[1]{\protect\rule{.1in}{.1in}}

\begin{document}
\preprint{gr-qc}
\title{A quantum black hole universe}

\author{M.B. Altaie}
\affiliation{Department of Physics, Yarmouk University, 21163 Irbid, Jordan}

\pacs{PACS number}

\begin{abstract}
Using the result obtained in a prevoius paper, in which I found an upper limit
on the region of particle creation in the vicinity of the event horizon of a
Schwarzschild black hole, and by assuming that all the created energy will be
absorbed by the black hole, a natural power law for the growth of the event
horizon is deduced. This result may explain the existence of galactic black
holes with very large masses. Application of this result on cosmological scale
shows that if we start with a Planck-sized black hole then the natural growth
of such a black hole will produce one with a density equals the present
critical density of the universe. Such a black hole universe will be in the
state of eternal inflation.

\end{abstract}

\date[Date text]{date}
\startpage{1}
\endpage{102}
\maketitle
\tableofcontents

\section{Introduction}

In a previous paper \cite{Altaie} I considered a heuristic derivation for
energy creation near the event horizon of a Schwarzschild black hole. The
derivation was very general and was based on arguments from the Heisenberg
uncertainty principle and the time dilation caused by the gravitational field
of the black hole. The result showed that there will be an upper limit for the
region in the vicinity of the event horizon within which such energy creation
may take place, this was given by%

\begin{equation}
R<\frac{4}{3}R_{s}\label{q22}%
\end{equation}
where $R_{s}=2M$ is the Schwarzschild radius.

Obviously such a proposal has nothing to do with the Hawking approach for
particle creation by black holes \cite{Hawking75}. However this kind of
argument can be related to the Casimir effect and the possibility for particle
creation by black holes through such mechanism.

According to the general theory of relativity the orbits of massive particles
near the event horizon of a Schwarzschild black hole are unstable below the
limit of $\sqrt{3}R_{s}$ (see for example \cite{WMT}), therefore all created
particles will eventually fall into the singularity through the event horizon.
The fate of all photons or any other massless particle will be the same.
Therefore, we can confidently assume that all created energy within the region
$R_{s}<R<\frac{4}{3}R_{s}$ will be absorbed by the black hole. This may
explain why we have no observational verification for the Hawking radiation,
despite long years of monitoring.

In this paper I am going to introduce the Casimir effect as a possible
mechanism to create particles in the vicinity of the event horizon. Such
phenomena has been studied by many authors \cite{Setaltaie} and was
extensively analyzed. In the next section I will elaborate how the Casimir
system is naturally constructed and will deduce a power law for the growth of
the surface area. In sec. (III) I show how a system of concentric Cauchy
surfaces can be constructed by the presence of the event horizon and the
barriers at distances with base of $\frac{4}{3}.$ In Sec. (IV) I will try to
model an inflating black hole universe based on results obtained from the
finite temperature correction to the vacuum energy in an Einstein universe,
where it was found that the non-zero vacuum energy renders the universe to
have a non-singular start.

\section{Particle Creation versus Casimir effect}

In flat spacetime Casimir \cite{Casimir} found that the vacuum fluctuations of
the electromagnetic field would give rise to an attractive force between two
parallel conducting flat plates, a negative energy density inversely
proportional to the fourth power of the distance between the two plates is
created as soon as such plates are introduced in the vacuum. This was called
the Casimir effect. Application of this effect in closed spacetimes (e.g. the
Einstein universe) has shown that it lead to a positive energy density (for
example see ref.\cite{Ford1}, \cite{Altaie1}). Further consideration of the
problem at finite temperatures resulted in calculating the finite temperature
corrections which was shown to be an important driver for the thermal
development of the universe when considered as a source for the Einstein field
equations \cite{Hu1}-\cite{Altaie2}.

Following Nugayev \cite{Nug1}, we may exchange the non-rotating black hole
with two spherical conductors one just outside the event horizon but very near
to it, and the other to be consider just below the upper regional limit of
$\frac{4}{3}R_{s}\,$. These two surfaces will constitute concentric shells,
analogous to the parallel conducting plates. Any amount of energy created
through the Casimir mechanism between the two shells is assumed to be added to
the total energy of the black hole, hence extending the event horizon by an
amount that has to be controlled by the conditions defined for the system.
This assumption seems acceptable in the light of the finding of Berezin et al.
\cite{Berezin1} that particles are created in pairs of positive energy by the
black hole, where one is emitted to infinity and the other falls on the black
hole causing a change of the inner structure. Then, we immediately deduce that
the new event horizon will have a radius of $\frac{4}{3}R_{s}$, and so the
growth of the event horizon will go on. This means that%

\begin{equation}
R_{n}=\left(  \frac{4}{3}\right)  ^{n}R_{s},\label{q23}%
\end{equation}
where $n=1,2,3....$.

The argument we place for considering discrete eigenvalues for the radius of
the event horizon is simple and goes as follows: The first Casimir system
which is composed of the event horizon and the barrier surface at $R_{1}%
=\frac{4}{3}R_{s}$ will create an amount of positive energy (the Casimir
energy) once formed will be added to the total energy of the black hole by the
argument of instability of orbits within the specified region. Therefore, the
event horizon will be extended to a new position and the second surface of the
new Casimir system (the new position of the barrier) will then be at
$R_{2}=\frac{4}{3}R_{1}$ and so it goes.

According to the above mechanism, the surface area of the event horizon of the
black holes will grow as%

\begin{equation}
A_{n}=\left(  \frac{4}{3}\right)  ^{2n}A_{s}.\label{q24}%
\end{equation}

This means that the area of the event horizon is quantized. The quantization
law here is much different from the law deduced by Bekenstein and Mukhanov
\cite{Bekenstein} according to which the spectrum of the surface area of the
event horizon was uniformly spaced. However, using the loop representation of
quantum gravity, Barreira et.al. \cite{barria} have shown that the
Bekenstein-Mukhanov area quantization spectrum is unrecoverable, consequently
they deduce that the Bekenstein-Mukhanov result is likely to be an artefact of
the ansatz used rather than a robust result.

From (\ref{q23}) it is clear that the expansion of the black hole will be
logarithmic (i.e. inflationary). The number of folding is given by%

\begin{equation}
n=8\log\left(  \frac{R_{n}}{R_{0}}\right) \label{q25a}%
\end{equation}

\section{A Constant-Time Hypersurface Structure}

Using the result in (\ref{q23}) above, we can construct a hypothetical
concentric Cauchy surface structure centered at the black hole singularity.
This structure is characterized by the eigenvalues of (\ref{q24}) and a set of
time-like Killing vectors normal to the surfaces. The transition from one
surface to another is associated with translation in time, and consequently
generation of energy. Therefore, each surface will represent an energy level
characterizing a black hole of the corresponding mass. The relative temporal
separation between these surfaces is given by the basic relation%

\begin{equation}
\frac{t(x_{1})}{t(x_{2})}=\left[  \frac{g_{00}(x_{1})}{g_{00}(x_{2})}\right]
^{-1/2}.\label{q25}%
\end{equation}

It is clear that the temporal separations between surfaces situated near to
$R_{s}$ are larger than those far away. If $x_{2}$ is a point at the
asymptotically-flat region where $t(x_{2})\equiv t_{\infty},$then we can write%

\begin{align}
t(x_{n})  &  =t_{\infty}[g_{00}(x_{n})]^{-1/2}\nonumber\\
&  =t_{\infty}\left[  1-\left(  \frac{3}{4}\right)  ^{n}\right]
^{-1/2}.\label{q26}%
\end{align}

This structure represents an infinite set of hypothetical Cauchy concentric
spherical shells surrounding the black hole prior to the development of the
black hole's event horizon. However, if the positive energy created by the
black hole within the specified region is to be added to the black hole's
total energy, then the mass development will take the following form%

\begin{equation}
M_{n}<\left(  \frac{4}{3}\right)  ^{n}M_{0},\label{q27}%
\end{equation}
where $M_{0}$ is the initial mass of the black hole.

Energy levels of the above structure have the following separations%

\begin{equation}
E_{n+1}-E_{n}<\frac{1}{3}\left(  \frac{4}{3}\right)  ^{n}E_{0},\label{q28}%
\end{equation}
where $E_{0}$ is the total initial mass-energy of the black hole.

From (\ref{q23}) and (\ref{q27}) we find that the energy density of the system
will develop according to the inequality%

\begin{equation}
\rho_{n}>\left(  \frac{3}{4}\right)  ^{2n}\rho_{0},\label{q29}%
\end{equation}
where $\rho_{0}$ is the initial density.

A system developing according to (\ref{q29}) would resemble an Einstein static
universe where the matter content of the universe is directly associated with
the radius of the spacial section. This can be seen once we relate (\ref{q23})
and (\ref{q29}) from which we get%
\begin{equation}
\frac{\rho_{n}}{\rho_{0}}\sim\frac{R_{0}^{2}}{R^{2}},\label{q29a}%
\end{equation}
where we have used the symbol $R_{0}$ instead of $R_{s}$ as trivial
substitution. This result motivates us to utilize the results obtained from
calculations performed for the Einstein universe as starting parameters for an
inflating black hole model for the universe.

\section{Cosmological Applications}

If the universe was born as a singularity, then it is unknown how it has
crossed its own event horizon. Although quantum effects may dissipate the
creation singularity, the presently available calculations, which incorporate
quantum fields into the classical curvature, do not indicate the possibility
of a universe born with crossed event horizon.

The available calculations \cite{Altaie1},\cite{Hu1}.\cite{Altaie2} indicates
that the universe was born as a finite-sized patch with a size less than the
Schwarzschild radius. This implies that the universe may have been born as a
black hole and still is. This idea is not new, and there are a number of
investigations that support it; for example, it was already shown long ago by
Oppenheimer and Snyder \cite{OS} that the interior of the Schwarzschild
solution could be described by a Friedmann universe. Moreover it was shown by
Pathria \cite{Path1} that our present universe may be described as an internal
Schwarzschild solution if it has the critical energy density. More recent
investigations \cite{Frolov1} based on the assumption of the existence of a
limiting curvature have shown that the inside of a Schwarzschild black hole
can be attached to a de Sitter universe at some space-like junction which is
taken to represent a short transition layer. Other scenarios in which the
universe emerges from the interior of a black hole were also proposed (see
refs. \cite{Easson1}-\cite{Daghigh}).On the other hand a universe that has a
critical density can easily be shown to have a Schwrzschild radius of a black
hole with equivalent total mass.

We will now utilize the results of the previous section to construct a model
for the whole universe. The model adopts the results of previous calculations,
\cite{Altaie1} and \cite{Altaie2}, of the back-reaction of the finite
temperature corrections to the vacuum energy density of the photon field in an
Einstein universe. Although the Einstein universe is static, the conformal
relation with the closed Robertson-Walker universe \cite{Ken} allow us \ to
consider the results as being of practical interest. In fact, the discrete
spectrum provided here by the inflating black hole model can be considered as
representing successions of different instantaneous states of the Einstein
static universe. The results of the back-reaction calculations have shown that
the thermal development of the universe covers two different regimes; the
Casimir regime, which extends over a very small range of the radius but huge
rise of temperature from zero to a maximum of $1.44\times10^{32}$K at a radius
of $5.5\times10^{-34}$cm. At this maximum temperature a phase transition takes
place, and the system crosses-over to the Planck regime, where photons get
emitted and absorbed freely exhibiting pure black-body spectrum. At this point
one can identify the primodial universe with an initial energy density
$\rho_{i}$ given by%

\begin{align}
\rho_{i}  &  =\alpha T^{4}\nonumber\\
&  =3.\,25\times10^{114}erg/cm^{3}.\label{q31}%
\end{align}

However, the same calculations \cite{Altaie2} showed that the Einstein
universe exhibits the presently measured microwave background temperature of
$2.73K$ at a radius of $1.83\times10^{30}cm.$ Consistency require us to adopt
this value for the radius of the present universe. Therefore, according to
(\ref{q29a}) we get%

\begin{equation}
\rho_{n}=2.\,92\times10^{-13}erg/cm^{3}.
\end{equation}

\ This result is very close to the radiation density in the present universe,
calculated in reference to the cosmic microwave background.

But one may argue that the estimated radius of the present universe (Hubble
length) is $1.38\times10^{28}cm$ and not $1.83\times10^{30}cm.$ Therefore
using (\ref{q29a}) again we get%

\begin{equation}
\rho_{now}^{\ast}\approx5.67\times10^{-30}gm.cm^{-3}.\label{q37}%
\end{equation}
a figure which is very close to the critical matter density which defines a
flat universe.

The above results are remarkable indeed and certainly implies some sort of
self-consistency on the side of the assumptions made in this work.

\section{Discussion and Conclusions}

The simple approach followed in this paper led us to a new quantization law
for the area of the event horizon, and consequently into a new area spectrum.
The main features of the new spectrum is that it is logarithmic and
macroscopic, in contrast to the Bekenstein-Mukhanov spectrum \cite{Bekenstein}%
, which was linear and microscopic (i.e. Planck dimensional). In fact the
Bekenstein-Mukhanov spectrum cannot be verified observationaly because in
practice the spectrum will look continuous for macroscopic black holes; no
fine structure details are measurable.

A logarithmic inflation arises naturally in our model as a result of the
Casimir system assumption. Normally such a model will benefit from all the
privileges of inflationary models over the standard big bang model, less their
conceptual problems. This is indeed the case since it was recently shown by
Easson and Brandenberger \cite{Easson1} that a universe born from the interior
of a black hole will not posses many of the problems of the standard big bang
model. In particular the horizon problem, the flatness problem and the problem
of formation of structures are solved naturally. This may well be the case for
a universe formed of the interior of an inflating black hole. Perhaps this is
the most important result that need to be analyzed further to see if one can
draw some observational consequences.

Our assumptions in this paper are strongly supported by the results we
obtained for the matter and radiation densities in the present universe. One
can see clearly that starting with a Planck-dimensional black hole universe,
the mechanism of the quantum development of such a hypothetical universe leads
to a universe having the present critical matter density, a point which is
supported by the recent observational investigations of the cosmic microwave
background radiation (see \cite{Bernardis} and \cite{Bennet}) . Further
analysis and development of this approach by investigating a Kerr or
Reissner-N\"{o}rdstrom black holes will be interesting, where one may expect
the emergence of a different scheme for area quantization.

One more feature that we remark in the present work is that our model exhibits
a discrete redshift for a time developing version. I mean that the relation in
(\ref{q23}) if used in the context of Hubble law for cosmic recession would
imply that velocities of the galactic clusters will appear to have discrete
redshift. Indeed the discrete redshift problem stands today as one
observational problem that is awaiting a solution \cite{Tiff}.

It might be of interest to point that Santilli \cite{Santilli} presented nine
theorems which marks inconsistencies in the general relativity theory (GRT).
The main point of interest in his contribution is his remark that GRT is
non-canonical at the clssical level and is non-unitary at the operator level.
This, it seems, is the main reason behind the explicit incompatibility of GRT
with quantum field theory (QFT). The need for a consistent approach may
require reformulating both GRT and QFT on some new common basis, and using may
be a new approach. The \emph{isogravitation} suggusted by Santilli may stand
to be a possible alternative. However, a more general approach to quantum
gavity may need to consider quantizing the gravitational field via an
alternative to the standard canonical quantization scheme, possibly through a
more profound mathematical approach which lead to comprehensive discreteness
and more explicit coherence of the physical world.

\end{document}